\documentclass[aps,pre,groupedaddress,showkeys,showpacs,amsmath,preprint,pre]{revtex4}

\usepackage{graphicx}

\begin{document}

\title{Microscopic formulation of the Zimm-Bragg model for the helix-coil transition}

\author{A.V.Badasyan}
\email{artem.badasyan@unive.it} 

\author{A.Giacometti}
\email{achille@unive.it} 

\affiliation{Dipartimento di Chimica Fisica,
Universita Ca' Foscari di Venezia,\\
Calle Larga S. Marta DD2137, I-30123 Venezia, Italy.
}

\author{Y.Sh.Mamasakhlisov}
\author{V.F.Morozov}
\email{morozov@ysu.am}
\affiliation{
Department of Molecular Physics, Yerevan State University,\\ A.Manougian Str.1, 375025, Yerevan, Armenia
}

\author{A.S.Benight}
\email{abenight@pdx.edu}
\affiliation{
Departments of Chemistry and Physics, Portland State University,\\ 1719 S.W. 10th Ave., Portland, OR 97207-0751, USA
}

\date{\today}

\begin{abstract}
A microscopic spin model is proposed for the phenomenological Zimm-Bragg model for the helix-coil transition in biopolymers.
 This model is shown to provide the same thermophysical properties of the original Zimm-Bragg
model and it allows a very convenient framework to compute statistical quantities. 
Physical origins of this spin model are made transparent by an exact mapping into a one-dimensional Ising
 model with an external field. However, the dependence on temperature of the reduced external field turns out to differ from
the standard one-dimensional Ising model and hence it gives rise to different thermophysical properties, despite the exact mapping
connecting them. We discuss how this point has been frequently overlooked in the recent literature. 
\end{abstract}


\maketitle
\begin{figure}[p]
\begin{center}
\includegraphics[scale=0.6]{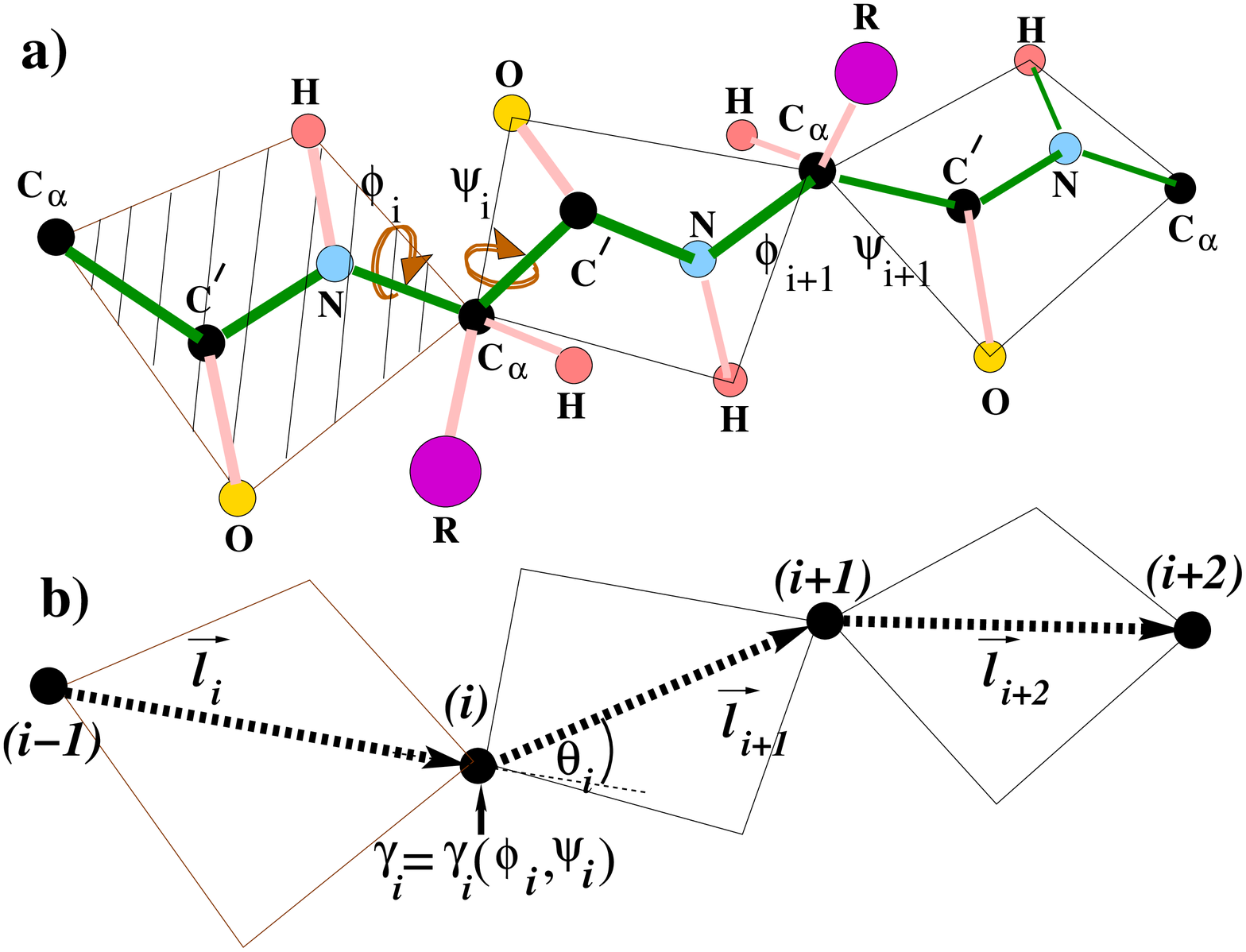}
\caption{\label{ff1} (COLOR ONLINE) A segment of a polypeptide chain in a \emph{trans} conformation is shown. Parallelograms indicate the plane of a virtual peptide bond. a) Diagramatic view of a polypeptide chain segment where the main-chain atoms are represented as rigid peptide units, linked by virtual bonds through the $C_\alpha$ atoms. Each unit has two degrees of freedom due to the rotation around the $C_\alpha-C^\prime$ (torsional angle $\phi$) and $N-C_\alpha$ (torsional angle $\psi$) bonds. $R$ stands for the amino acid residues, all other atoms have corresponding chemical labels aside. b) Coarse-grained representation of a polypeptide chain: the configuration is described with the help of fictitious vectors $\vec{l_i}$, that depend on coordinates of two neighboring $C_\alpha$ atoms and bond angle $\theta_i=\pi-\arccos{\frac{ \vec{l}_i \vec{l}_{i+1} } { l_i l_{i+1} }}$, that depends on coordinates of three carbons, and a pair of torsional angles $\phi_i,\:\psi_i$.
}
\end{center}
\end{figure}

Statistical descriptions of polypeptide chain conformations involve important coarse-graining on the level of the $C_\alpha$ atoms.
Due to planarity of the amide group, torsional angles of successive repeated units can be considered independent, 
and a pair of $\phi_i, \psi_i$ angles can be associated with each repeated unit (see Fig.~\ref{ff1}(a)). 
Introduction of virtual bonds, connecting neighboring $C_\alpha$'s, strongly simplifies the description. 
Thus the configuration of a polypeptide chain can be described with the array of bond vectors $\{\vec{l}_i\}$, $i=1...N-1$, 
related to its backbone, and a new variable $\gamma_i$ representing the state of the particular 
$i-$th units (Fig.~\ref{ff1}(b))\cite{molbiol,flory}. Within the framework of the helix-coil transition theory, this variable could have different values in the helix and coil states, taken according
to some particular prescription. 
Since Doty and co-workers experimentally showed that polypeptide chains in solution can be reversibly converted from the random coil to $\alpha$-helix conformations \cite{Doty56}, a number of methods have been proposed to model the phenomenon \cite{polsher} ranging from highly sophisticated computer simulations \cite{h-c_simul} to highly simplified spin-like models \cite{eaton}. In the past few decades significant advances have been made in computational capabilities of both computer hardware and software
thus enabling investigations to an unprecedented level of complexity. At the same time, interpretations of results from single-molecule techniques, such as stretching with optical tweezers \cite{stretch}, have largely relied on applications of classical spin models \cite{force_melt}. In addition these spin approaches remain attractive for describing folding of helical proteins \cite{folddill} and influences of solvent on secondary structure formation and stability \cite{pinc_02}. Among these, the Zimm-Bragg (ZB) model stands out for its success \cite{Zimm59}. While very useful in interpreting experimental results, the original ZB theory is far less satisfactory from a theoretical point of view as it lacks a well defined microscopic description, thus preventing a clear connection with more sophisticated levels of theoretical description.\par
It has been frequently stated in the literature, that the ZB model can be described, at a microscopic level, by a one-dimensional Ising model (see \emph{e.g.} Ref.\cite{wrong-ising} for recent representative examples). However, care must be exercised in making this assertion. The aim of this short note is directed toward formally addressing the actual equivalence of the ZB and one-dimensional Ising models.\par
The comparative analysis will be performed by introducing a one-dimensional Potts-like spin model and demonstrating that it gives rise to the same thermodynamic properties as the original ZB model. The Potts-like model is then mapped, via an exact transformation, into a one-dimensional Ising-like model with an external field. It is shown that due to the temperature dependence in the external field, predicted thermophysical properties are not equivalent to the standard one-dimensional Ising model.\par

In its simplest formulation the ZB model is based on the following combinatorial rules \cite{Zimm59}. 
(1) Every repeated unit exists in either the hydrogen-bonded (helical) or un-bonded (coil) state; 
(2) Every un-bonded repeated unit contributes a statistical weight of unity to the partition function;
(3) Every bonded repeated unit that follows another bonded repeated unit, contributes a statistical weight, $s$;
(4) Every bonded repeated unit that follows two or more un-bonded repeated units, contributes a statistical weight of $s\sigma$ 
(in conjunction with Rule 3, this defines $\sigma$);
(5) Every bonded repeated unit that follows less than two un-bonded repeated units, contributes a statistical weight of zero.\par
Note that $s$ has the meaning of a statistical weight, and is usually interpreted in terms of a free energy change $s=e^{- \beta (G_{helix}-G_{coil})}$ between the helix and coil states. Similar interpretation can be given for $\sigma$ with the additional restriction of being purely entropic in nature (unlike $s$ which has both enthalpic and entropic contributions) \cite{Takano02}.
Using combinatorial techniques, the ZB model allows derivation of the thermodynamic properties of the system from the eigenvalues 
of the following $2 \times 2$ matrix \cite{Zimm59}, 
\begin{equation}
\mathbf{M}_{ZB}= 
\begin{pmatrix}
1&1 \\ \sigma s &s
\end{pmatrix} 
\label{zb-matrix}
\end{equation}
 The corresponding secular equation providing the eigenvalues is,
\begin{equation}
\lambda^2-\lambda(s+1)+s(1-\sigma)=0,
\label{zb-sec-eq}
\end{equation}
which \textit{defines} the thermodynamics of the ZB model. Note that the matrix in Eq.~(\ref{zb-matrix}) is \textit{not} symmetric. This contrasts with the results obtained from the standard one-dimensional Ising model with an external field, which is symmetric \cite{Huang63}. As a result, eigenvalues, and hence the thermodynamics are different in the two cases. We shall return to this point later on where comparisons between the ZB model and one-dimensional Ising model with a particular external field will be discussed.\par
Our microscopic formulation of the ZB model is founded on a Potts-like formulation akin to the more general model, previously presented within a slightly different context \cite{moroz-biopoly}. Assume that spin $\gamma_i$ describing the state of the $i$-th repeated unit can take one of $Q(\geq 2)$ values; $\gamma_i=1$  corresponding to values of the torsional angles $\phi_i$ and $\psi_i$ from the helical region of the Ramachandran map, while the other $Q-1$ identical values correspond to torsional 
angles from the coil region. As we shall see, the condition $Q\ge 2$ plays a fundamental role as it accounts for the large degeneracy (and hence entropy) of the coil state. The magnitude of $Q$ can be identified with the ratio of the allowed region area versus helical region area on a Ramachandran map. Although the formation of one hydrogen bond fixes the values of three couples of torsional angles (see, \emph{e.g.} \cite{moroz-biopoly}), we simplify the model and consider only a nearest-neighbor construction. The energy of interaction is assumed to be different from zero when both $\gamma_i$ and $\gamma_{i+1}$ are equal to $1$. The corresponding spin Hamiltonian is, 
\begin{equation}
-\beta H=\beta J\sum\limits_{i=1}^{N}\delta(\gamma_i,1)\delta(\gamma_{i+1},1), 
\label{ham-basic}
\end{equation}
\noindent where $\delta$'s are Kronecker symbols and $\beta=1/\kappa_B T$ is the inverse thermal energy. Here and in the following, we consider open boundary conditions with the $N\rightarrow \infty$ limit taken at the end. We further remark that the Potts-like model defined in Eq.~(\ref{ham-basic}) differs from the classical Potts model proposed by Goldstein \cite{Goldstein84} for the helix-coil transition (a detailed comparison between the two models is interesting and will be addressed in future work).\par
The partition function $Z$ can be obtained via standard transfer matrix techniques \cite{fisher1,flory},
\begin{equation}
Z=\sum\limits_{\lbrace \gamma_i \rbrace} e^{-\beta H(\lbrace \gamma_{i} \rbrace)}=
\sum\limits_{\lbrace \gamma_i \rbrace} \prod\limits_{i=1}^{N} \left(\mathbf{M}\right)_{\gamma_i,\gamma_{i+1}},
\label{pfunc}
\end{equation}
where $(\mathbf{M})_{\gamma_i,\gamma_{i+1}}$ are the elements of the $Q \times Q$ matrix
\begin{equation}
\mathbf{M}(Q \times Q)=
\begin{pmatrix}
e^{\beta J} & 1 & ... & 1 \\ 
1 & 1 & ... & 1 \\
... & ...& ... & ...\\
1 & 1 & ... & 1 
\end{pmatrix} .
\label{pfunc2}
\end{equation}
From the structure of the matrix in Eq.~(\ref{pfunc2}), it is clear 
that there are only two linearly independent eigenvectors, so the number of nontrivial eigenvalues $\lambda$ and the order of the characteristic equation is also equal to two. This can be explicitly verified as follows. Write the characteristic equation $\lvert \mathbf{M}-\mathbf{I}\lambda \rvert=0$, $\mathbf{I}$ being the identity matrix. Successively subtract the second row from the first row, third row from the second, etc., until all rows have been accounted for. Similarly, successively add column $Q$ with column $Q-1$, $Q-1$ with $Q-2$ etc, so as to obtain a final block diagonal determinant with the characteristic equation,
\begin{equation}
\lambda^{Q-2} \times \det
\begin{pmatrix}
e^{\beta J}-1-\lambda & e^{\beta J}-1 \\ 
1 & Q-\lambda 
\end{pmatrix} 
=0.
\label{det-gmpc}
\end{equation}
\noindent Neglecting the $Q-2$ trivial eigenvalues, a simple change of variables $\Lambda = \lambda/Q$,
$\sigma=Q^{-1}$ and $s=(e^{\beta J}-1)/Q$ yields
\begin{equation}
\det
\begin{pmatrix}
s-\Lambda & s \\ 
\sigma & 1-\Lambda 
\end{pmatrix}
=\Lambda^2-\Lambda(s+1)+s(1-\sigma)=0,
\label{gmpc2-sec-eq}
\end{equation}
\noindent which exactly coincides with the characteristic equation for the ZB model given in Eq.~(\ref{zb-sec-eq}). Therefore, the Hamiltonian in Eq.~(\ref{ham-basic}) provides exactly the same thermodynamics of the ZB model and, hence, can be considered equivalent to it. Next, the relationship between the ZB and Ising models is examined. The partition function in Eq.~(\ref{pfunc}) of the Hamiltonian in Eq.~(\ref{ham-basic}) can be cast in the following form,
\begin{equation}
Z=\sum\limits_{\gamma_1=1}^Q \sum\limits_{\gamma_2=1}^Q e^{\beta J\delta(\gamma_1,1)\delta(\gamma_{2},1)} \sum\limits_{\gamma_3=1}^Q 
e^{\beta J\delta(\gamma_2,1)\delta(\gamma_{3},1)}... \sum\limits_{\gamma_N=1}^Q e^{\beta J\delta(\gamma_{N-1},1)\delta(\gamma_{N},1)}.
\label{trick1}
\end{equation}
\noindent After each of the sums, a term $\sum\limits_{m_i=0,1} \delta(m_i,\delta(\gamma_i,1))$ can be inserted, since it is equal to unity. Upon changing the summation order and tracing out over $\gamma$ variables, one immediately gets the partition function, 
\begin{equation}
Z=\sum\limits_{ \lbrace m_i \rbrace } \prod\limits_{i=1}^{N} \exp{[ \beta J m_i m_{i+1}+q(1-m_i)]}=\sum\limits_{ \lbrace m_i \rbrace }\exp{[-\beta H]},
\label{trick2}
\end{equation}
\noindent where the Hamiltonian is given by,
\begin{equation}
-\beta H_{zb}=K_{zb}\sum_{i}m_i m_{i+1}+\mu_{zb}\sum_{i}(1-m_i).
\label{trick-ham}
\end{equation}
\noindent Here the coupling $K_{zb}=\beta J$ represents the reduced energy of a hydrogen bond and $\mu_{zb}=q=\ln{(Q-1)}$ plays the role of a reduced chemical potential within a lattice gas formulation. For comparison, the Hamiltonian of a one-dimensional Ising model in a similar lattice gas formulation is given by,
\begin{equation}
- \beta H_{Ising} = K_{Ising} \sum_{i} m_i m_{i+1} + \mu_{Ising} \sum_{i}(1-m_i), 
\label{ising}
\end{equation}
\noindent where $K_{Ising}=\beta J=K_{zb}$, and $\mu_{Ising}= \beta q $. The last term on the right in Eq.~(\ref{ising}), $\mu_{Ising}$, is temperature dependent, whereas the analogous term on the right in (\ref{trick-ham}), $\mu_{zb}$, is not. Herein constitutes a fundamental difference between the ZB and one-dimensional Ising models.\par
\begin{figure}[p]
\begin{center}
\includegraphics[scale=0.5]{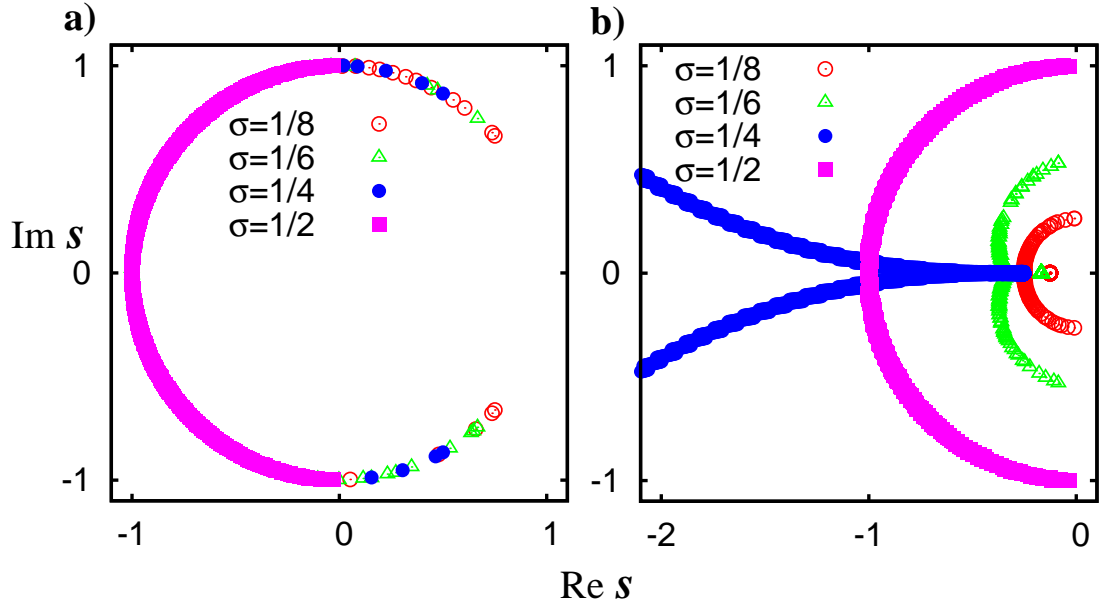}
\caption{\label{ff5} (COLOR ONLINE) Temperature (Fisher) zeros in terms of the $s=(e^{\beta J}-1)/(e^q-1)$ at different values of parameter $\sigma=1/Q=(e^q-1)^{-1}$ for: a) Zimm-Bragg model given by Eq.~(\ref{ham-basic}) and b) Ising-like model, given by Eq.~(\ref{ising}). As obvious, Fisher zeroes for the $\sigma=1/2$ case are similar for both models, since it corresponds to the model with $\mu_{ising}= \mu_{zb} =0 $ field.
}
\end{center}
\end{figure}

 
Corroboration comes from considering the zeroes of the partition functions \cite{polsher} of the ZB model (as defined by the Hamiltonian in Eq.~(\ref{ham-basic}), or its equivalent Hamiltonian in Eq.~(\ref{trick-ham})) and comparing the zeros with those obtained for the partition function for the one-dimensional Ising Hamiltonian in Eq.~(\ref{ising}). The method of zeroes of partition functions is a standard tool for identifying phase transitions in spin models \cite{yanglee}. As the control variable is temperature, the zeroes must be considered in the complex temperature plane, known as the Fisher zeroes \cite{fisher2}. In the thermodynamic limit the point where zeroes cross the real positive axis can be identified as the transition point. We remark, that in the absence of long-range interactions the helix-coil transition is not a true phase transition even for infinite polymer lengths ($N \rightarrow \infty$), and hence, no crossing of the real axis should be expected. On the other hand, recent numerical studies of Fisher zeroes for helix-coil models with long-range interactions, and finite chain lengths, strongly suggest a first order transition \cite{hansman}.  For short-range interactions and infinite chains, such as the case treated here, the distributions of Fisher zeroes are clearly different and have been discussed by Poland and Sheraga \cite{polsher} for some limiting values of parameters. Their calculation is repeated here within a more general framework and results are compared with those obtained from the corresponding one-dimensional Ising model \cite{fisher2}. These different scenarios are a direct consequence of the different temperature dependence mentioned above.  The Fisher zeros are depicted in Figure~\ref{ff5}(a) in the $\text{Re}\:s-\text{Im}\:s$ plane for different values of the $\sigma$ parameter. Here $s=(e^{\beta J}-1)/(e^q-1)$ and $\sigma=1/(e^{q}-1)$ are the most convenient variables to perform this comparison. Poland and Sheraga in \cite{polsher} have shown that at limiting values of $\sigma$, Fisher zeroes of the ZB model lie on the unit circle, so do the edge zeroes. Expected scaling of the edge zeroes with $\sigma$ was predicted to be $\text{Re}(s_{\text{edge}})=1-2\sigma; \;\text{Im}(s_{\text{edge}})=2\sigma^{1/2}(1-\sigma)^{1/2}$, so that the phase transition limit is approached as $\sigma$ decreases, only reaching it in the $\sigma \rightarrow 0$ limit of infinite cooperativity. Within the microscopic formulation given by Eq.~(\ref{trick-ham}), we have performed a numerical check and confirmed the scaling exactly as above for the vast range of $\sigma$ values. Thus, the real part of the edge zero indicates the transition temperature and the imaginary part can be
considered as a measure of the cooperativity or of correlations present in the system at the transition point \cite{ananik}. Edge zeroes of the Ising model (Fig.~\ref{ff5}(b)), as given by Eq.~(\ref{ising}), do not lie on the circle, clearly indicating the difference in scaling with $\sigma$ values. Both the cooperativities and the stabilities clearly differ in the two cases. While the transition point for the ZB model is close to $s=1$, it lies in the vicinity of $s=0$ for the Ising model, as shown in Fig.~\ref{ff5}(b). As previously anticipated, a numerical simulation study of a three-dimensional model with long-range interactions have shown that Fisher zeroes occupy the same spherical region of unit radius, as in Fig.~\ref{ff5}(a), but nearly cross the positive real axis unlike the short-ranged ZB counterpart \cite{hansman}.\par
In conclusion, we have argued that a proper microscopic description of the Zimm-Bragg model is not a standard one-dimensional Ising model, as often tacitly assumed in the recent
literature, but rather a one-dimensional Potts-like model. This can indeed be shown to be
equivalent to an Ising-like one-dimensional model having however different properties with
respect to the usual Ising counterpart, as made evident from a comparative Fisher zeroes analysis.

\begin{acknowledgments}
Authors would like to acknowledge Amos Maritan, Marco Zamparo, Ruben Ghughazaryan and Vadim Ohanyan for insightful discussions and 
comments and thank Amos Maritan for the idea of transformation from Eq.~(\ref{trick1}) to Eq.~(\ref{trick2}). AB and AG acknowledge the support from PRIN-COFIN 2007 grant.
\end{acknowledgments}



{}

\newpage


\newpage

\end{document}